# Dynamics of multiple bubbles, excited by femtosecond filament in water: Role of aberrations

F.V.Potemkin[*], E.I. Mareev

*Faculty of Physics and International Laser Center M.V. Lomonosov Moscow State University, Leninskie Gory, bld.1/62, 119991, Moscow, Russia*
*Corresponding author: potemkin@automationlabs.ru*

**Abstract:** Using shadow photography, we observed microsecond time scale evolution of multiple cavitation bubbles, excited by tighty focused femtosecond laser pulse in water under supercritical power regime (~100 $P_{cr}$). In these extreme conditions high energy delivery into the microvolume of liquid sample leads to creation of single filament which becomes a source of cavitation region formation. When aberrations were added to the optical scheme the hot spots along the filament axis are formed. At high energies (more than 40μJ) filaments in these hot spots are fired and, as a result, complex pattern of cavitation bubbles is created. The bubbles can be isolated from each other or build exotic "drop-shaped" cavitation region, which evolution at the end of its "life", before the final collapse, contains the jet emission. The dynamics of the cavitation pattern was investigated from pulse energy and focusing. We found that greater numerical aperture of the focusing optics leads to greater cavitation area length. The strong interaction between bubbles was also observed. This leads to a significant change of bubble evolution, which is not yet in accordance with Rayleigh model.

The phenomenon of laser-induced cavitation is well known [1], but still attracts attention to itself due to a complexity of processes, accompanying cavitation bubble evolution. The cavitation bubble rapid growth and collapse can cause shock waves, rapid jets and sonoluminescence [2]. The aggressive nature of laser-induced cavitation has found a broad range of applications, such as cell lysis [3], cell membrane poration [4] and ocular surgery [5]. Cavitation is used for mixing, pumping, switching and moving objects in microfluids. The laser-induced bubbles are used for creation of highly focused supersonic microjets. Due to their excellent controllability, high velocity and relatively low power requirements, these jets are an attractive option for needle-free drug injection [6].

Tight focusing of ultrashort laser pulses with even microjoule pulse energy into the bulk of transparent dielectric leads to extreme intensities (>$10^{13}$W/cm$^2$) in microvolume and plasma formation. The mean energy of plasma electrons reaches the value of several electronvolts. The initial plasma distribution dramatically depends on laser pulse parameters and focusing geometry [7]. Due to high teperatures the thin layer of vapor is generated, which consequently transformed into cavitation bubble. The shape of laser-induced cavitation bubble depends on the plasma density distribution and, thus, the initial spatial profile of laser intensity in the medium defines the spatio-temporal evolution of cavitation bubble. The Rayleigh model (or its more correct modifications) describes the dynamics of isolated laser-induced cavitation bubble $\rho R \ddot{R} + (3/2)\rho \dot{R}^2 = p_i - p_e$, where $R$ – the radius of cavitation bubble, $p_i$ – the pressure inside the bubble, $p_e$ – outer pressure [1,7]. The model also can be applied to cylindrical or semi-spherical cavitation bubbles with some simple modifications [8]. But when two or more bubbles begin to interact with each other, or when the shape is strongly non-spherical, the cavitation bubbles evolution considerably differs from one, described by Rayleigh model [6,9,10]. Thus, the additional investigations should be performed in this field.

Nowadays the research on thermo- and hydrodynamics of laser filament on the microsecond time scale is gaining popularity [11–15]. Many scientific groups are concentrated on studying femtosecond laser filament evolution in gases. But there is a lack of works which represent the experimental results on the dynamics of post effects, induced by femtosecond filament, which, in turn, was fired by tightly focused beam under supercritical power regime, in condensed matter on the microsecond timescales. Only in liquids and porous media the cavitation bubble formation is possible, that give us the opportunity to estimate the energy delivered to the medium [16].

In this Letter we report the first, from our knowledge, complex study of the dynamics of multiple cavitation bubbles, induced by a single femtosecond filament in water, as a convenient prototype of condensed matter, under supercritical power regime. The supercritical regime and tight focusing geometry enable effective energy delivery to the medium, which leads to generation of cavitation bubbles pattern with high controllability in their parameters. Moreover, we focus in our study on the role of aberrations in cavitation bubbles generation and evolution.

To observe the dynamics of cavitation bubbles on the microsecond timescales the shadow photography technique was applied. The experimental setup is sketched at Fig.1. As a probe pulse the second harmonic of Nd:YAG nanosecond laser (wavelength is 532 nm, pulse duration 10 ns, repetition rate is 10 Hz) was used. The time delay between two lasers was controlled electronically by the delay generator in 80 microseconds delay range with 10ns step.

The cavitation bubbles were excited by a femtosecond filament, which was produced by a Cr:Forsterite femtosecond pulse (wavelength is 1240 nm, pulse duration about 140 fs, intensity contrast is about $10^4$, repetition rate is 10Hz). The energy of the pump pulse can be varied up to 190μJ which corresponds to peak power of 900 MW (taking into account linear absorption in water), which is approximately one hundred times above the critical power for water. Pump laser pulse was focused into the water cell, creating plasma, shock waves and sequence of cavitation bubbles. The experiments were carried out with two sets of focusing parameters: tight focusing with numerical aperture NA=0.4 and focal length of 3.3mm (lens L1), and relatively loose focusing – numerical aperture NA= 0.3 and focal length of 8mm (lens L2). Turning the focusing lenses on 180 degrees with respect to laser radiation allowed us to artificially add the spherical aberrations into the scheme. In order to characterize focusing conditions the experiments were carried out in air and in water. When aspheric lens L1 was placed in air a bright spark with 4 μm in diameter and about 30 μm in length was observed. In water instead a plasma channel (length more than 100 μm) was formed. To avoid the aberrations in other cases the aspheric lens was placed inside the water cell.

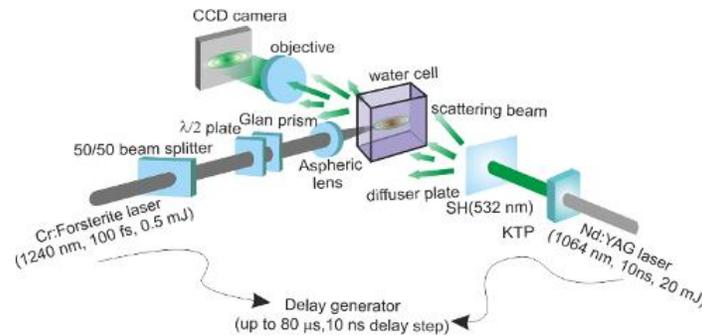

FIG.1. Experimental setup. The pump pulse from Cr:Forsterite femtosecond laser was focused into the water cell. The energy of pump pulse was varied by half-wave plate and Glan prism. The probe pulse was a radiation of the second harmonic of Nd:YAG laser, passed through the water cell with cavitation bubble. It was collected by the microscope objective on the CCD camera. The bubble modifies refractive index, which leads to the dark areas on the CCD matrix. The time delay between optical pulses can be electronically tuned up to 80 microseconds with 10ns step.

When an ultrashort laser pulse with peak power much above the critical power propagates through media, a dynamical balance between Kerr self-focusing, plasma defocusing and diffraction occurs, and laser filament is fired. Instead of collimated and loosely focused beams in tight focusing geometry the laser intensity in the focal spot can reach a value up to $10^{15}$ W/cm$^2$ (it is an upper limit of a rough estimate) [17–19]. Accordingly the electron density can reach $0,1 n_{cr}$ ($n_{cr}=m_e\omega^2/4\pi e^2 \cong 7.3\times 10^{20}$ cm$^{-3}$) and the energy of plasma electrons is sufficiently larger than in other cases [20]. In our conditions the energy is localized in the microvolume with 4 μm in diameter. As a result only single filament can be formed. To determine the contribution of processes taking place during filament propagation (plasma defocusing, Kerr self-focusing and diffraction) the simple estimates can be made. The length of self-focusing can be calculated as $L_{sf}=\lambda/2\pi n_2 I \approx 1$μm ($n_2=1,6\times 10^{-16}$ cm$^2$/W), the length scale for plasma defocusing $L_{defoc}=L_{pl}n_{at}/n_e=n_0\lambda n_{cr}/\pi n_e \approx 5$μm and the diffraction length is a Rayleigh length, which is about 15 μm [21]. Therefore the Kerr self-focusing don`t allow the laser radiation to leave the optical axis and one continuous filament with approximately uniform distribution of electron concentration along the optical axis can be formed [22]. This can be confirmed by the fact that the radii of cavitation bubbles are uniform along the filament axis, due to their proportionality to plasma electrons energy (see Fig.2). When enough energy dissipates through linear and non-linear absorption the filament stops.

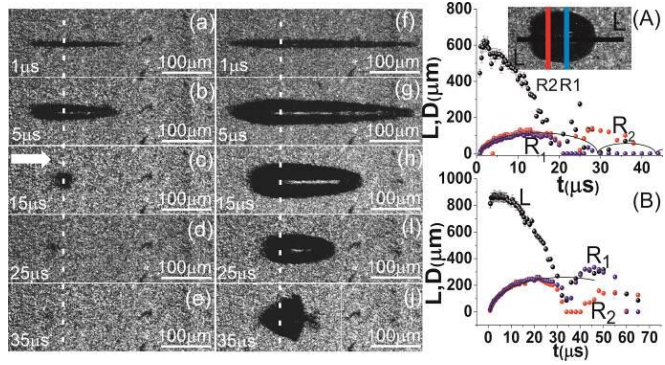

FIG.2. The evolution of cavitation bubbles excited by femtosecond filament in water at different energies (38±4μJ in the left column and (A), 190±10μJ in the right column and (B)). The focal plane position is given by dashed white line. The laser pulse direction and time delay between pump and probe pulses are shown in the figures. The scale bar is 100 μm. At (A),(B) the diameter and the length of the cavitation are shown. The sold line is the approximation based on Rayleigh model. Lines in the inset picture demonstrate where the diameters and length of the cavitation area were measured.

From the shadow photographs the cavitation area length can be measured. Just after formation the cavitation area replicates the plasma distribution, therefore the filament length could be restored from shadow photographs (see Fig.2). In our experiments we compared the results obtained with different focusing lenses. We found that greater focal length leads to greater filament length. For example, when the lens L1 was used, the filament length was about 1.5 times greater than in the case of the lens L2. This result is at odds with conventional notions of powerful laser radiation filamentation for collimated or loosely focused beams [21]. We suppose that the main two processes lead to the filament growth with the increase of the focusing sharpness. The first process is linear absorption (the distance between the lens and geometrical focus is greater in the case of lens L2). The second process is the intensity decrease in the focal plane of lens L2 in comparison with the lens L1, therefore the Kerr self-focusing is less and can`t efficiently keep the energy near optical axis. The measurements of the filament length showed that it has the logarithmic dependence on laser pulse energy (see Fig.3a) which is in a good agreement with results in [23,24].

When the aberrations were added to optical scheme, cavitation bubbles are formed in hot spots, which are located in aberration focuses. If the laser radiation was focused by lens L2, cavitation bubbles in hot spots are sufficiently separated from each other (see Fig.4). These bubbles are initially spherical, but with the energy increase their shape becomes more close to cylindrical one. It is related to the fact, that with the energy increase the bubble grows outward laser pulse direction. The length of each cavitation bubble depends logarithmically on laser energy (see Fig.3b). Therefore the laser filament is fired in each hot spot. At high energies (more than 40μJ) intensity clamping starts to play the main role and the radii of neighboring bubbles flatten out. Aberrations lead to the laser energy redistribution over the hot spots along the optical axis and, as a result, the intensity in the first hot spot is decreased. When more tight focusing geometry is used (see Fig.5.) the contribution of the Kerr self-focusing is less than diffraction, therefore filament can be fired only in first bubbles at high intensities (more than $10^{13}W/cm^2$). In other cases the cavitation bubbles have a spherical shape.

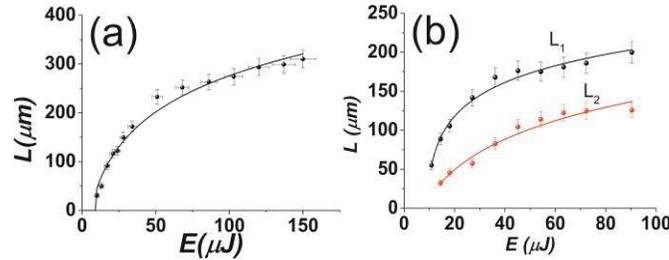

FIG.3 (a) The filament length L and (b) the length of the filaments in aberration hot spots as a function of laser energy E. Line corresponds to logarithmic dependence.

Now let us concentrate on the cavitation region evolution. The initial cavitation region shape replicates the spatial distribution of laser-induced plasma. Then the cavitation region rapidly grows. The growth of the cavitation region radius is in a good agreement with the Rayleigh model. After reaching the maximal size it starts to collapse (see Fig.2). In that moment the pressure inside the cavitation region is much smaller than the outer pressure [1].

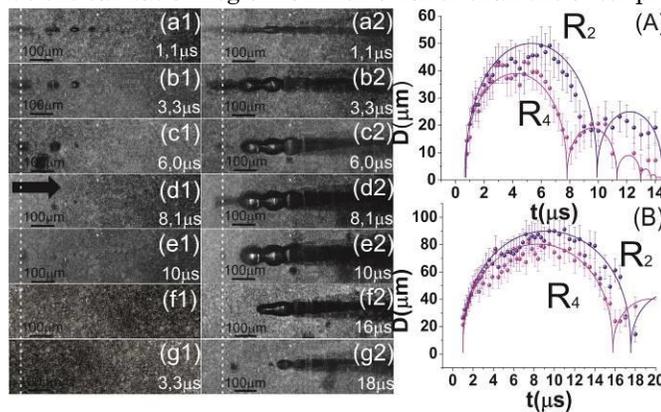

FIG.4. The evolution of cavitation bubbles induced by focused femtosecond pulse in water (lens L2 with aberrations was used). The energy of laser pulse: (a1)-(e1),(A) -30±4μJ, (a2)-(g2),(B) - 190±10μJ. The focal plane position is given by dashed white line. The laser pulse direction and time delay between pump and probe pulses are shown in the figures. The scale bar is 100 μm. At (A), (B) the diameter of the second and fourth cavitation bubble is shown. The solid line is the approximation based on Rayleigh model.

The collapse of the cavitation region induced by laser radiation focused by lens L2 starts at the outer bubbles (see Fig.2,4). The speed of collapse along the optical axis is larger than in the perpendicular direction. The rapid collapse of outer bubbles also leads to energy flux forward to the center of cavitation region. This flux is symmetric about the center of cavitation region. Therefore the cavitation bubbles collapses to the small area in the center of initial cavitation region.

When single continuous filament is formed the bubbles are completely overlapped (see Fig.2). Overlapping leads to energy exchange between cavitation bubbles, the exchange becomes stronger, when the difference between bubble diameters (energies and temperatures) grows. The evolution of the cavitation region in the case when aberrations were added to optical scheme is in a good agreement with a Rayleigh model, despite of energy exchange between cavitation bubbles (see Fig.4 A,B). This can be easily explained using the fact, that the cavitation bubbles have approximately uniform diameter and energy, therefore the exchange is insignificant. At small energies when aberrations were added to optical scheme the cavitation bubbles are isolated from each other and the energy exchange between them is negligible (see Fig.4a1). When a single filament is formed, only the growth of cavitation region diameter can be described by the Rayleigh model. The process of cavitation region collapse sufficiently different from one predicted by the Rayleigh model, because the processes of energy exchange between bubbles dramatically change the evolution of the region. But when cavitation bubbles grow, the pressure and energy gradients are directed from the bubble, and therefore forces, acting on the bubble boundary, do not sufficiently differ from forces arisen in the case of spherical bubble growth. The attempt of approximation of cavitation area evolution, assuming its shape as a cylinder, give much worse result than in the case of the spherical shape of the bubbles. Accordingly, the cavitation area can be considered as a superposition of overlapped spherical bubbles, which interact with each other.

For lens L1 all cavitation bubbles are spherical and the radius of each consequent bubble is less, than the previous one (Fig. 5a). The evolution of cavitation bubbles is in good agreement with the Rayleigh model for laser energy below $14\pm2\mu J$ and for bubbles distant from the first "hot spot", because in the first order of approximation such bubbles are spherical and their interaction with each other can be neglected. Bubbles begin to overlap at energies above $14\pm2\mu J$. The overlapping leads to energy exchange between the bubbles and the discrepancy from the Rayleigh model is taken place (see Fig.5A). At the long time delays (more than 2 microseconds) and pump energies above $80\pm6\mu J$ (see Fig.5) the bubbles are completely overlapped and indistinguishable from each other. At time delays more than 3μs they form one "drop-shaped" cavitation region, which length is greater than the maximal diameter. The radii of first bubbles are approximately equal, therefore they have the equal energy. Due to this fact the energy exchange between the first bubble and its neighbors is insignificant.

When the "drop" collapses, we have a significant difference in energy and pressure gradient on the opposite sides of the structure (see Fig.5 d-f), which is similar to the bubble collapse in the case of lens L2. Thus, the length of the cavitation region decreases much greater than the diameter (see Fig.5 A). Such rapid decrease of the length leads to energy and momentum flux that directed to the center of the cavitation area. The smaller bubbles collapse first, therefore, the flux directed toward the laser radiation and the jets appear (see Fig.5 g-j). After averaging shadow photographs over twenty pictures, we also found, that the average shape of the cavitation region is a modified mirror image of the initial bubbles pattern, because the energy flux drives the energy towards the laser, which leads to secondary cavitation bubble formation before the geometric focus. The conservation of the cavitation bubble shape on the microsecond time scale still makes possible to determine initial spatial plasma density distribution. One can say that cavitation "remembers" initial energy distribution on the filament axis up to microsecond time scales.

FIG.5 The evolution of cavitation bubbles induced by focused ( lens L1 with aberrations was used) femtosecond pulse (energy is 190±10μJ) in water. At (A) the energy is 30±4μJ, The focal point is given by white line. The laser pulse direction and time delay between pump and probe pulses are shown on the shadow photographs. The scale bar is 100 μm (a)-(d) the growth of cavitation bubble, (c)-(j) its collapse. At the (h)-(j) jets formation is shown. At (A),(B) the L is the length of cavitation region and $R_i$ – diameter of i-th bubble. Lines shows approximation based on the Rayleigh model.

In conclusion, in tight focusing geometry with peak power of laser radiation much above the critical power a single filament can be formed. If aberrations are added to optical scheme filaments will be fired in aberration focuses. Overlapping of the filaments can efficiently increase the length of cavitation area. The filament length logarithmically depends on laser energy. The filaments become the centers of cavitation bubble formation. At different regimes of filament formation cavitation bubbles can be isolated from each other, form one cylindrical cavitation region or construct one "drop-shaped" cavitation region. We found that bubbles evolution is completely described by the Rayleigh model when cavitation bubbles are completely isolated from each other. The overlapping causes the energy exchange between the bubbles. The influence of the exchange on every bubble formation depends on the energy difference between neighbor bubbles. When the radii (and energies) of the bubbles are sufficiently different from each other the Rayleigh model could not still be applied to that bubbles behavior description. During the collapse the outer bubbles collapse first and transfer their energy to inner bubbles or to the surrounding medium, emitting jets.

This research has been supported by the Russian Foundation for Basic Research (Project No. 14-02-00819a) and partly by the M.V. Lomonosov Moscow State University Program of Development.

[